\documentclass[twoside,12pt]{article}
\usepackage{epsfig}

\newcommand{\be}{\begin{equation}}
\newcommand{\ee}{\end{equation}}
\newcommand{\bea}{\begin{eqnarray}}
\newcommand{\eea}{\end{eqnarray}}

\begin{document}

\title{ \vspace{1cm} Neutrinos in cosmology}
\author{S.\ Hannestad$^{1}$\\
\\
$^1$Department of Physics and Astronomy, University of Aarhus,
\\ Ny Munkegade, DK-8000 Aarhus C, Denmark} \maketitle
\begin{abstract}
I review the basics of neutrino cosmology, from the question of
neutrino decoupling and the presence of sterile neutrinos to the
effects of neutrinos on the cosmic microwave background and large
scale structure. Particular emphasis is put on cosmological
neutrino mass measurements, both the present bounds and the future
prospects.
\end{abstract}

\section{Introduction} 

Neutrinos are the second most abundant particles in our Universe.
This means they have a profound impact on many different aspects
of cosmology, from the question of leptogenesis in the very early
universe, over big bang nucleosynthesis, to late time structure
formation. Here I review some general apects of neutrino cosmology
with particular emphasis on issues relevant to cosmological bounds
on the neutrino mass.

The absolute value of neutrino masses are very difficult to
measure experimentally. On the other hand, mass differences
between the light neutrino mass eigenstates, $(m_1,m_2,m_3)$, can
be measured in neutrino oscillation experiments.

The combination of all currently available data suggests two
important mass differences in the neutrino mass hierarchy. The
solar mass difference of $\delta m_{12}^2 \simeq 7.1-8.9 \times
10^{-5}$ eV$^2 \,\, (3\sigma)$ and the atmospheric mass difference
$\delta m_{23}^2 \simeq 1.4-3.3 \times 10^{-3}$ eV$^2  \,\,
(3\sigma)$
\cite{Maltoni:2004ei,Maltoni:2003da,Aliani:2003ns,deHolanda:2003nj}
(see \cite{Valle:2005ai} for a recent review).

In the simplest case where neutrino masses are hierarchical these
results suggest that $m_1 \sim 0$, $m_2 \sim \delta m_{\rm
solar}$, and $m_3 \sim \delta m_{\rm atmospheric}$. If the
hierarchy is inverted one instead finds $m_3 \sim 0$, $m_2 \sim
\delta m_{\rm atmospheric}$, and $m_1 \sim \delta m_{\rm
atmospheric}$. However, it is also possible that neutrino masses
are degenerate, $m_1 \sim m_2 \sim m_3 \gg \delta m_{\rm
atmospheric}$, in which case oscillation experiments are not
useful for determining the absolute mass scale.

Experiments which rely on kinematical effects of the neutrino mass
offer the strongest probe of this overall mass scale. Tritium
decay measurements have been able to put an upper limit on the
electron neutrino mass of 2.3 eV (95\% conf.) \cite{kraus}.
However, cosmology at present yields an much stronger limit which
is also based on the kinematics of neutrino mass.

Very interestingly there is also a claim of direct detection of
neutrinoless double beta decay in the Heidelberg-Moscow experiment
\cite{Klapdor-Kleingrothaus:2001ke,Klapdor-Kleingrothaus:2004wj},
corresponding to an effective neutrino mass in the $0.1-0.9$ eV
range. If this result is confirmed then it shows that neutrino
masses are almost degenerate and well within reach of cosmological
detection in the near future.

Another important question which can be answered by cosmological
observations is how large the total neutrino energy density is.
Apart from the standard model prediction of three light neutrinos,
such energy density can be either in the form of additional,
sterile neutrino degrees of freedom, or a non-zero neutrino
chemical potential.

In section 2 I review the present cosmological data which can be
used for analysis of neutrino physics. In section 3 I discuss
neutrino physics around the epoch of neutrino decoupling at a
temperature of roughly 1 MeV, including the relation between
neutrinos and Big Bang nucleosynthesis. Section 4 discusses
neutrinos as dark matter particles, including mass constraints on
light neutrinos, and sterile neutrino dark matter. Finally,
section 5 contains a discussion.

\section{Cosmological data} 

\paragraph{Large Scale Structure (LSS) --}

At present there are two large galaxy surveys of comparable size,
the Sloan Digital Sky Survey (SDSS)
\cite{Tegmark:2003uf,Tegmark:2003ud} and the 2dFGRS (2~degree
Field Galaxy Redshift Survey) \cite{2dFGRS}. The SDSS is still
ongoing, but will be completed very soon. It will then be
significantly larger than the 2dF. Furthermore, a continuation in
the form of a new and extended survey, SDSS-II, has already been
approved (see http://www.sdss.org).

\paragraph{Cosmic Microwave Background (CMB) --}

The CMB temperature fluctuations are conveniently described in
terms of the spherical harmonics power spectrum $C_l^{TT} \equiv
\langle |a_{lm}|^2 \rangle$, where $\frac{\Delta T}{T}
(\theta,\phi) = \sum_{lm} a_{lm}Y_{lm}(\theta,\phi)$.  Since
Thomson scattering polarizes light, there are also power spectra
coming from the polarization. The polarization can be divided into
a curl-free ($(E)$) and a curl ($(B)$) component, much in the same
way as $\vec{E}$ and $\vec{B}$ in electrodynamics can be derived
from the gradient of a scalar field and the curl of a vector field
respectively (see for instance \cite{Kamionkowski:1996ks} for a
very detailed treatment). The polarization introduced a sequence
of new power spectra, but because of different parity some of them
are explicitly zero. Altogether there are four independent power
spectra: $C_l^{TT}$, $C_l^{EE}$, $C_l^{BB}$, and the $T$-$E$
cross-correlation $C_l^{TE}$.

The WMAP experiment has reported data only on $C_l^{TT}$ and
$C_l^{TE}$ as described in
Refs.~\cite{Bennett:2003bz,Spergel:2003cb}. Other experiments,
while less precise in the measurement of the temperature
anisotropy and not providing full-sky coverage, are much more
sensitive to small scale anisotropies and to CMB polarization.
Particularly the ground based CBI \cite{Pearson:2002tr}, DASI
\cite{Kovac:2002fg}, and ACBAR \cite{Kuo:2002ua} experiments, as
well as the BOOMERANG balloon experiment
\cite{Jones:2005yb,Piacentini:2005yq,Montroy:2005yx} have provided
useful data.

\paragraph{Type Ia supernovae}

Observations of distant supernovae have been carried out on a
large scale for about a decade. In 1998 two different projects
almost simultaneously published measurements of about 50 distant
type Ia supernovae, out to a redshift or about 0.8
\cite{Riess:1998cb,Perlmutter:1998np}. These measurements were
instrumental for the measurement of the late time expansion rate
of the universe.

Since then a, new supernovae have continuously been added to the
sample, with the Riess et al. \cite{Riess:2004} "gold" data set of
157 distant supernovae being the most recent. This includes
several supernovae measured by the Hubble Space Telescope out to a
redshift of 1.7.

Very recently, the first data has been released from the Supernova
Legacy Survey (SNLS) \cite{Astier:2005qq}, providing the currently
largest data set of Type Ia supernovae.

\paragraph{Other data --}

Apart from CMB, LSS and SNI-a data there are a number of other
cosmological measurements of importance to neutrino cosmology.
Perhaps the most important is the measurement of the Hubble
constant by the HST Hubble Key Project, $H_0=72 \pm 8~{\rm
km}~{\rm s}^{-1}~{\rm Mpc}^{-1}$ \cite{Freedman:2000cf}.

\section{Neutrino Decoupling} 

\subsection{Standard model}

In the standard model neutrinos interact via weak interactions
with $e^+$ and $e^-$. In the absence of oscillations neutrino
decoupling can be followed via the Boltzmann equation for the
single particle distribution function \cite{kolb}
\begin{equation}
\frac{\partial f}{\partial t} - H p \frac{\partial f}{\partial p}
= C_{\rm coll}, \label{eq:boltz}
\end{equation}
where $C_{\rm coll}$ represents all elastic and inelastic
interactions. In the standard model all these interactions are $2
\leftrightarrow 2$ interactions in which case the collision
integral for process $i$ can be written
\begin{eqnarray}
C_{\rm coll,i} (f_1) & = & \frac{1}{2E_1} \int \frac{d^3 {\bf
p}_2}{2E_2 (2\pi)^3} \frac{d^3 {\bf p}_3}{2E_3 (2\pi)^3} \frac{d^3
{\bf p}_4}{2E_4 (2\pi)^3} \nonumber \\
&& \,\, \times (2\pi)^4 \delta^4
(p_1+p_2-p_3+p_4)\Lambda(f_1,f_2,f_3,f_4) S |M|^2_{12 \to 34,i},
\end{eqnarray}
where $S |M|^2_{12 \to 34,i}$ is the spin-summed and averaged
matrix element including the symmetry factor $S=1/2$ if there are
identical particles in initial or final states. The phase-space
factor is $\Lambda(f_1,f_2,f_3,f_4) = f_3 f_4 (1-f_1)(1-f_2) - f_1
f_2 (1-f_3)(1-f_4)$.

The matrix elements for all relevant processes can for instance be
found in Ref.~\cite{Hannestad:1995rs}. If Maxwell-Boltzmann
statistics is used for all particles, and neutrinos are assumed to
be in complete scattering equilbrium so that they can be
represented by a single temperature, then the collision integral
can be integrated to yield the average annihilation rate for a
neutrino
\begin{equation}
\Gamma = \frac{16 G_F^2}{\pi^3} (g_L^2 + g_R^2) T^5,
\end{equation}
where
\begin{equation}
g_L^2 + g_R^2 = \cases{\sin^4 \theta_W + (\frac{1}{2}+\sin^2
\theta_W)^2 & for $\nu_e$ \cr \sin^4 \theta_W +
(-\frac{1}{2}+\sin^2 \theta_W)^2 & for $\nu_{\mu,\tau}$}.
\end{equation}

This rate can then be compared with the Hubble expansion rate
\begin{equation}
H = 1.66 g_*^{1/2} \frac{T^2}{M_{\rm Pl}}
\end{equation}

 to
find the decoupling temperature from the criterion $\left. H =
\Gamma \right|_{T=T_D}$. From this one finds that $T_D(\nu_e)
\simeq 2.4$ MeV, $T_D(\nu_{\mu,\tau}) \simeq 3.7$ MeV, when $g_*
=10.75$, as is the case in the standard model.

This means that neutrinos decouple at a temperature which is
significantly higher than the electron mass. When $e^+e^-$
annihilation occurs around $T \sim m_e/3$, the neutrino
temperature is unaffected whereas the photon temperature is heated
by a factor $(11/4)^{1/3}$. The relation $T_\nu/T_\gamma =
(4/11)^{1/3} \simeq 0.71$ holds to a precision of roughly one
percent. The main correction comes from a slight heating of
neutrinos by $e^+e^-$ annihilation, as well as finite temperature
QED effects on the photon propagator
\cite{Dicus:1982bz,Rana:1991xk,herrera,Dolgov:1992qg,Dodelson:1992km,%
Fields:1993zb,Hannestad:1995rs,Dolgov:1997mb,Dolgov:1999sf,gnedin,%
Esposito:2000hi,Steigman:2001px,Mangano:2001iu,osc,Mangano:2005cc}.

\subsection{Big Bang nucleosynthesis and the number of neutrino species}

Shortly after neutrino decoupling the weak interactions which keep
neutrons and protons in statistical equilibrium freeze out. Again
the criterion $\left. H = \Gamma \right|_{T=T_{\rm freeze}}$ can
be applied to find that $T_{\rm freeze} \simeq 0.5 g_*^{1/6}$ MeV
\cite{kolb}.

Eventually, at a temperature of roughly 0.2 MeV deuterium starts
to form, and very quickly all free neutrons are processed into
$^4$He. The final helium abundance is therefore roughly given by
\begin{equation}
Y_P \simeq \left. \frac{2 n_n/n_p}{1+n_n/n_p} \right|_{T\simeq 0.2
\,\, {\rm MeV}}.
\end{equation}

$n_n/n_p$ is determined by its value at freeze out, roughly by the
condition that $n_n/n_p|_{T=T_{\rm freeze}} \sim
e^{-(m_n-m_p)/T_{\rm freeze}}$.

Since the freeze-out temperature is determined by $g_*$ this in
turn means that $g_*$ can be inferred from a measurement of the
helium abundance. However, since $Y_P$ is a function of both
$\Omega_b h^2$ and $g_*$ it is necessary to use other measurements
to constrain $\Omega_b h^2$ in order to find a bound on $g_*$. One
customary method for doing this has been to use measurements of
primordial deuterium to infer $\Omega_b h^2$ and from that
calculate a bound on $g_*$. Usually such bounds are expressed in
terms of the equivalent number of neutrino species, $N_\nu \equiv
\rho/\rho_{\nu_0}$, instead of $g_*$. The exact value of the bound
is quite uncertain because there are different and inconsistent
measurements of the primordial helium abundance (see for instance
Ref.~\cite{Barger:2003zg} for a discussion of this issue). The
most recent analyses are \cite{Barger:2003zg} where a value of
$1.7 \leq N_\nu \leq 3.0$ (95\% C.L.) was found,
\cite{Serpico:2004gx} which found $-1.14 \leq \Delta N_\nu \leq
0.73$, and \cite{Cyburt:2004yc} which found that $N_\nu =
3.14^{+0.7}_{-0.65}$ at 68\% C.L. The difference in these results
can be attributed to different assumptions about uncertainties in
the primordial helium abundance.

Another interesting parameter which can be constrained by the same
argument is the neutrino chemical potential, $\xi_\nu=\mu_\nu/T$
\cite{Kang:xa,Kohri:1996ke,Orito:2002hf,Ichikawa:2002vn}. At first
sight this looks like it is completely equivalent to constraining
$N_\nu$. However, this is not true because a chemical potential
for electron neutrinos directly influences the $n-p$ conversion
rate. Therefore the bound on $\xi_{\nu_e}$ from BBN alone is
relatively stringent ($-0.1 \leq \xi_{\nu_e} \leq 1$
\cite{Kang:xa}) compared to that for muon and tau neutrinos
($\left|\xi_{\nu_{\mu,\tau}}\right| < 7$ \cite{Kang:xa}). However,
as will be seen in the next section, neutrino oscillations have
the effect of almost equilibrating the neutrino chemical
potentials prior to BBN, completely changing this conclusion.

\subsection{The number of neutrino species - joint CMB and BBN analysis}

The BBN bound on the number of neutrino species presented in the
previous section can be complemented by a similar bound from
observations of the CMB and large scale structure. The CMB depends
on $N_\nu$ mainly because of the early Integrated Sachs Wolfe
effect which increases fluctuation power at scales slightly larger
than the first acoustic peak. The large scale structure spectrum
depends on $N_\nu$ because the scale of matter-radiation equality
is changed by varying $N_\nu$.

Many recent papers have used CMB, LSS, and SNI-a data to calculate
bounds bounds on $N_\nu$
\cite{Crotty:2003th,Hannestad:2003xv,Pierpaoli:2003kw,Barger:2003zg,%
Cuoco:2003cu}, and some of the bounds are listed in Table
\ref{table:nnu}. Recent analyses combining BBN, CMB, and large
scale structure data can be found in
\cite{Hannestad:2003xv,Barger:2003zg}, and these results are also
listed in Table \ref{table:nnu}.

Common for all the bounds is that $N_\nu=0$ is ruled out by both
BBN and CMB/LSS. This has the important consequence that the
cosmological neutrino background has been positively detected, not
only during the BBN epoch, but also much later, during structure
formation.

The most recent bound which uses the SNI-a "gold" data set, as
well as the new Boomerang CMB data finds a limit of $N_\nu =
4.2^{+1.7}_{-1.2}$ at 95\% C.L. The bound from late-time
observations is now as good as that from BBN, and the two derived
value are mutually consistent given the systematic uncertainties
in the primordial helium value.

In Fig.~1 we show the currently allowed region for $\Omega_b h^2$
and $N_\nu$ (taken from \cite{Hannestad:2005jj}).

\begin{figure}[htbp]
\begin{center}
\epsfig{file=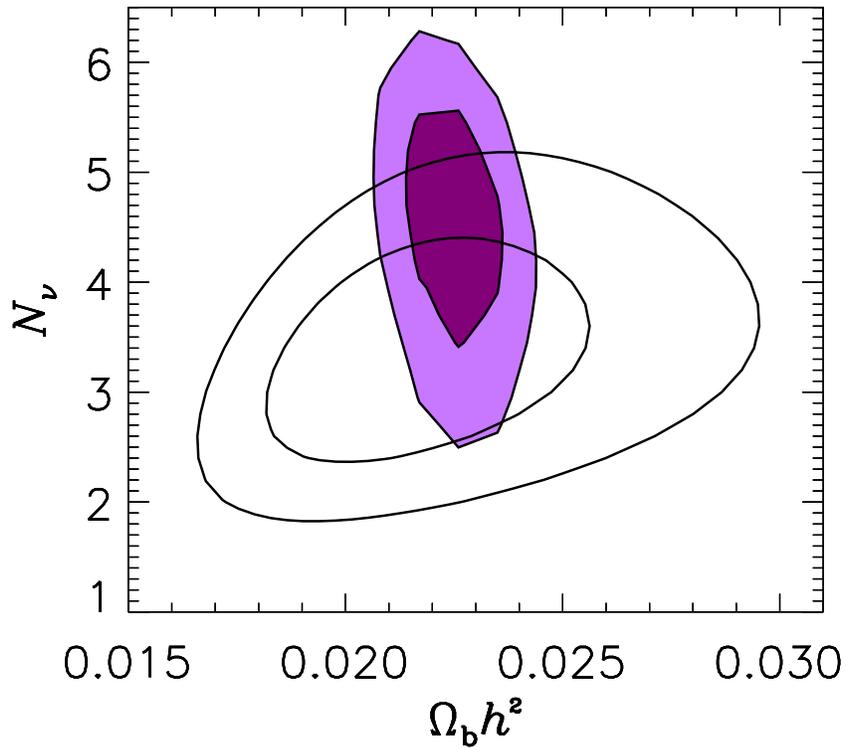,width=0.6\textwidth}
\end{center}
\bigskip
\caption{ The 68\% (dark) and 95\% (light) likelihood contours for
$\Omega_b h^2$ and $N_\nu$ for all available data. The other
contours are 68\% and 95\% regions for BBN, assuming the $^4$He
and D values given in \protect\cite{Cyburt:2004yc}.}
\end{figure}

\begin{table}
\begin{center}
\caption{Various recent limits on the effective number of neutrino
species, as well as the data used.}
\begin{tabular}{@{}lll}
\hline

Ref. & Bound on $N_\nu$ & Data used \\
\hline

Crotty et al. \cite{Crotty:2003th} & $1.4 \leq N_\nu \leq 6.8$ &
CMB, LSS \\

Hannestad \cite{Hannestad:2003xv} & $0.9 \leq N_\nu \leq 7.0$ &
CMB, LSS \\

Pierpaoli \cite{Pierpaoli:2003kw} & $1.9 \leq N_\nu \leq 6.62$ &
CMB, LSS \\

Barger et al. \cite{Barger:2003zg} & $0.9 \leq N_\nu \leq 8.3$ &
CMB \\

Hannestad \cite{Hannestad:2005jj} & $3.0 \leq N_\nu \leq 5.9$ &
CMB, LSS, SNI-a \\

\hline
\end{tabular}
\end{center}
\label{table:nnu}
\end{table}

\subsection{The effect of oscillations}

In the previous section the one-particle distribution function,
$f$, was used to describe neutrino evolution. However, for
neutrinos the mass eigenstates are not equivalent to the flavour
eigenstates because neutrinos are mixed. Therefore the evolution
of the neutrino ensemble is not in general described by the three
scalar functions, $f_i$, but rather by the evolution of the
neutrino density matrix, $\rho \equiv \psi \psi^\dagger$, the
diagonal elements of which correspond to $f_i$.

For three-neutrino oscillations the formalism is quite
complicated. However, the difference in $\Delta m_{12}$ and
$\Delta m_{23}$, as well as the fact that $\sin 2 \theta_{13} \ll
1$ means that the problem effectively reduces to a $2 \times 2$
oscillation problem in the standard model. A detailed account of
the physics of neutrino oscillations in the early universe is
outside the scope of the present paper, however an excellent and
very thorough review can be found in Ref.~\cite{dolgov}

Without oscillations it is possible to compensate a very large
chemical potential for muon and/or tau neutrinos with a small,
negative electron neutrino chemical potential \cite{Kang:xa}.
However, since neutrinos are almost maximally mixed a chemical
potential in one flavour can be shared with other flavours, and
the end result is that during BBN all three flavours have almost
equal chemical potential. This in turn means that the bound on
$\nu_e$ applies to all species so that
\cite{lunardini,Pastor:2001iu,Dolgov:2002ab,Abazajian:2002qx,Wong:2002fa}.

\begin{equation}
|\xi_i| = \frac{|\eta_i|}{T} < 0.15
\end{equation}
for $i=e,\mu,\tau$.

In models where sterile neutrinos are present even more remarkable
oscillation phenomena can occur. However, I do not discuss this
possibility further, except for the possibility of sterile
neutrino warm dark matter, and instead refer to the review
\cite{dolgov}.

\subsection{Low reheating temperature and neutrinos}

In most models of inflation the universe enters the normal,
radiation dominated epoch at a reheating temperature, $T_{\rm
RH}$, which is of order the electroweak scale or higher. However,
in principle it is possible that this reheating temperature is
much lower, of order MeV. This possibility has been studied many
times in the literature, and a very general bound of $T_{\rm RH}
\geq 1$ MeV has been found
\cite{Kawasaki:1999na,Kawasaki:2000en,Giudice:2000ex,Giudice:2000dp}

This very conservative bound comes from the fact that the light
element abundances produced by big bang nucleosynthesis disagree
with observations if the universe if matter dominated during BBN.
However, a somewhat more stringent bound can be obtained by
looking at neutrino thermalization during reheating. If a scalar
particle is responsible for reheating then direct decay to
neutrinos is suppressed because of the necessary helicity flip.
This means that if the reheating temperature is too low neutrinos
never thermalize. If this is the case then BBN predicts the wrong
light element abundances. However, even if the heavy particle has
a significant branching ratio into neutrinos there are problems
with BBN. The reason is that neutrinos produced in decays are born
with energies which are much higher than thermal. If the reheating
temperature is too low then a population of high energy neutrinos
will remain and also lead to conflict with observed light element
abundances. A recent analysis showed that in general the reheating
temperature cannot be below roughly 4 MeV \cite{Hannestad:2004px}.

\section{Neutrino Dark Matter} 

Neutrinos are a source of dark matter in the present day universe
simply because they contribute to $\Omega_m$. The present
temperature of massless standard model neutrinos is $T_{\nu,0} =
1.95 \, K = 1.7 \times 10^{-4}$ eV, and any neutrino with $m \gg
T_{\nu,0}$ behaves like a standard non-relativistic dark matter
particle.

The present contribution to the matter density of $N_\nu$ neutrino
species with standard weak interactions is given by
\begin{equation}
\Omega_\nu h^2 = N_\nu \frac{m_\nu}{92.5 \, {\rm eV}}
\end{equation}
Just from demanding that $\Omega_\nu \leq 1$ one finds the bound
\cite{Gershtein:gg,Cowsik:gh}
\begin{equation}
m_\nu < \frac{46 \, {\rm eV}}{N_\nu} \label{eq:mnu}
\end{equation}

\subsection{The Tremaine-Gunn bound}

If neutrinos are the main source of dark matter, then they must
also make up most of the galactic dark matter. However, neutrinos
can only cluster in galaxies via energy loss due to gravitational
relaxation since they do not suffer inelastic collisions. In
distribution function language this corresponds to phase mixing of
the distribution function \cite{Tremaine:we}. By using the theorem
that the phase-mixed or coarse grained distribution function must
explicitly take values smaller than the maximum of the original
distribution function one arrives at the condition
\begin{equation}
f_{\rm CG} \leq f_{\nu,{\rm max}} = \frac{1}{2}
\end{equation}
Because of this upper bound it is impossible to squeeze neutrino
dark matter beyond a certain limit \cite{Tremaine:we}. For the
Milky Way this means that the neutrino mass must be larger than
roughly 25 eV {\it if} neutrinos make up the dark matter. For
irregular dwarf galaxies this limit increases to 100-300 eV
\cite{Madsen:mz,salucci}, and means that standard model neutrinos
cannot make up a dominant fraction of the dark matter. This bound
is generally known as the Tremaine-Gunn bound.

Note that this phase space argument is a purely classical
argument, it is not related to the Pauli blocking principle for
fermions (although, by using the Pauli principle $f_\nu \leq 1$
one would arrive at a similar, but slightly weaker limit for
neutrinos). In fact the Tremaine-Gunn bound works even for bosons
if applied in a statistical sense \cite{Madsen:mz}, because even
though there is no upper bound on the fine grained distribution
function, only a very small number of particles reside at low
momenta (unless there is a condensate). Therefore, although the
exact value of the limit is model dependent, limit applies to any
species that was once in thermal equilibrium. A notable
counterexample is non-thermal axion dark matter which is produced
directly into a condensate.

\subsection{Neutrino hot dark matter}

A much stronger upper bound on the neutrino mass than the one in
Eq.~(\ref{eq:mnu}) can be derived by noticing that the thermal
history of neutrinos is very different from that of a WIMP because
the neutrino only becomes non-relativistic very late.

In an inhomogeneous universe the Boltzmann equation for a
collisionless species is \cite{MB}
\begin{equation}
L[f] = \frac{Df}{D\tau} = \frac{\partial f}{\partial \tau}
 + \frac{dx^i}{d\tau}\frac{\partial f}{\partial x^i} +
\frac{dq^i}{d\tau}\frac{\partial f}{\partial q^i} = 0,
\end{equation}
where $\tau$ is conformal time, $d \tau = dt/a$, and $q^i = a p^i$
is comoving momentum. The second term on the right-hand side has
to do with the velocity of the distribution in a given spatial
point and the third term is the cosmological momentum redshift.

Following Ma and Bertschinger \cite{MB} this can be rewritten as
an equation for $\Psi$, the perturbed part of $f$
\begin{equation}
f(x^i,q^i,\tau) = f_0(q) \left[ 1 + \Psi(x^i,q^i,\tau) \right]
\end{equation}

In synchronous gauge that equation is

\begin{equation}
\frac{1}{f_0}[f] = \frac{\partial \Psi}{\partial \tau} + i
\frac{q}{\epsilon} \mu \Psi + \frac{d \ln f_0}{d \ln q}
\left[\dot{\eta}-\frac{\dot{h}+6\dot{\eta}} {2} \mu^2 \right] =
\frac{1}{f_0} C[f],
\end{equation}
where $q^j = q n^j$, $\mu \equiv n^j \hat{k}_j$, and $\epsilon =
(q^2 + a^2 m^2)^{1/2}$. $k^j$ is the comoving wavevector. $h$ and
$\eta$ are the metric perturbations, defined from the perturbed
space-time metric in synchronous gauge \cite{MB}
\begin{equation}
ds^2 = a^2(\tau) [-d\tau^2 + (\delta_{ij} + h_{ij})dx^i dx^j],
\end{equation}
\begin{equation}
h_{ij} = \int d^3 k e^{i \vec{k}\cdot\vec{x}}\left(\hat{k}_i
\hat{k}_j h(\vec{k},\tau) +(\hat{k}_i \hat{k}_j - \frac{1}{3}
\delta_{ij}) 6 \eta (\vec{k},\tau) \right).
\end{equation}

Expanding this in Legendre polynomials one arrives at a set of
hierarchy equations
\begin{eqnarray}
\dot{\delta} & = & -\frac{4}{3} \theta - \frac{2}{3} \dot h \nonumber \\
\dot{\theta} & = & k^2\left(\frac{\delta}{4} - \sigma \right) \nonumber \\
2 \dot{\sigma} & = & \frac{8}{15} \theta - \frac{3}{15} k F_3
+ \frac{4}{15} \dot h + \frac{8}{5} \dot{\eta} \nonumber \\
\dot{F}_l & = & \frac{k}{2l+1} \left(l F_{l-1} - (l+1) F_{l+1}
\right)
\end{eqnarray}
For subhorizon scales ($\dot h = \dot \eta = 0$) this reduces to
the form
\begin{eqnarray}
\dot{\delta} & = & -\frac{4}{3} \theta \nonumber \\
\dot{\theta} & = & k^2\left(\frac{\delta}{4} - \sigma \right) \nonumber \\
2 \dot{\sigma} & = & \frac{8}{15} \theta - \frac{3}{15} k F_3 \nonumber \\
\dot{F}_l & = & \frac{k}{2l+1} \left(l F_{l-1} - (l+1) F_{l+1}
\right)
\end{eqnarray}

One should notice the similarity between this set of equations and
the evolution hierarchy for spherical Bessel functions. Indeed the
exact solution to the hierarchy is
\begin{equation}
F_{l}(k \tau) \sim j_l(k \tau)
\end{equation}
This shows that the solution for $\delta$ is an exponentially
damped oscillation. On small scales, $k > \tau$, perturbations are
erased.

This in intuitively understandable in terms of free-streaming.
Indeed the Bessel function solution comes from the fact that
neutrinos are considered massless. In the limit of CDM the
evolution hierarchy is truncated by the fact that $\theta=0$, so
that the CDM perturbation equation is simply $\dot \delta = -\dot
h/2$. For massless particles the free-streaming length is $\lambda
= c \tau$ which is reflected in the solution to the Boltzmann
hierarchy. Of course the solution only applies when neutrinos are
strictly massless. Once $T \sim m$ there is a smooth transition to
the CDM solution. Therefore the final solution can be separated
into two parts: 1) $k > \tau(T=m)$: Neutrino perturbations are
exponentially damped 2) $k < \tau(T=m)$: Neutrino perturbations
follow the CDM perturbations. Calculating the free streaming
wavenumber in a flat CDM cosmology leads to the simple numerical
relation (applicable only for $T_{\rm eq} \gg m \gg T_0$)
\cite{kolb}

\begin{equation}
\lambda_{\rm FS} \sim \frac{20~{\rm Mpc}}{\Omega_x h^2}
\left(\frac{T_x}{T_\nu}\right)^4 \left[1+\log \left(3.9
\frac{\Omega_x h^2}{\Omega_m h^2} \left(\frac{T_\nu}{T_x}\right)^2
\right)\right]\,. \label{eq:freestream}
\end{equation}

\begin{figure}[htbp]
\begin{center}
\epsfig{file=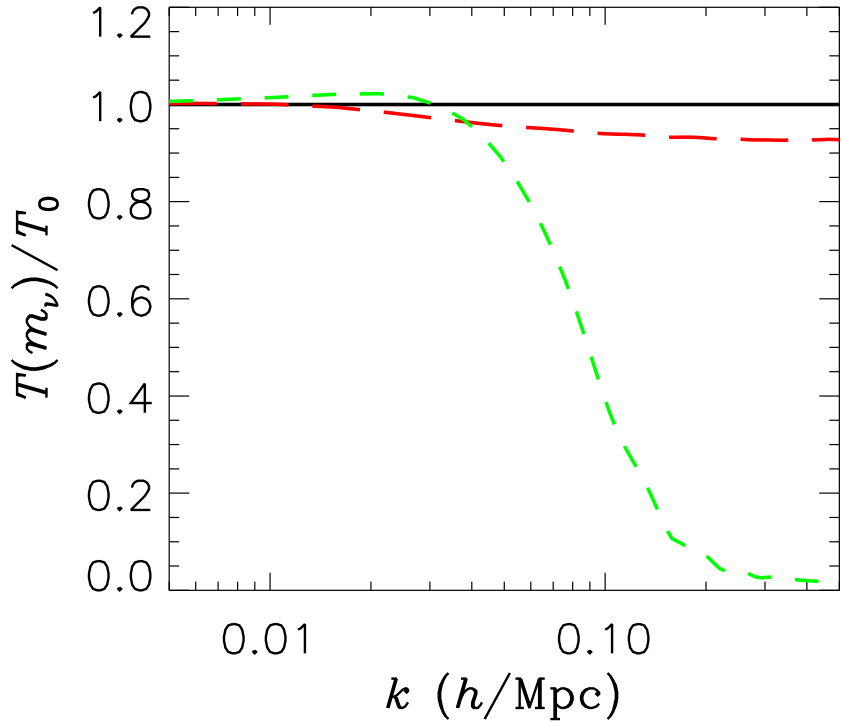,width=0.6\textwidth}
\end{center}
\bigskip
\caption{\label{fig:nutrans} The transfer function $T(k,t=t_0)$
for various different neutrino masses. The solid (black) line is
for $m_\nu=0$, the long-dashed for $m_\nu = 0.3$ eV, and the
dashed for $m_\nu=1$ eV.}
\end{figure}

In Fig.~\ref{fig:nutrans} I have plotted transfer functions for
various different neutrino masses in a flat $\Lambda$CDM universe
$(\Omega_m+\Omega_\nu+\Omega_\Lambda=1)$. The parameters used were
$\Omega_b = 0.04$, $\Omega_{\rm CDM} = 0.26 - \Omega_\nu$,
$\Omega_\Lambda = 0.7$, $h = 0.7$, and $n=1$.

When measuring fluctuations it is customary to use the power
spectrum, $P(k,\tau)$, defined as
\begin{equation}
P(k,\tau) = |\delta|^2(\tau).
\end{equation}
The power spectrum can be decomposed into a primordial part,
$P_0(k)$, and a transfer function $T(k,\tau)$,
\begin{equation}
P(k,\tau) = P_0(k) T(k,\tau).
\end{equation}
The transfer function at a particular time is found by solving the
Boltzmann equation for $\delta(\tau)$.

At scales much smaller than the free-streaming scale the present
matter power spectrum is suppressed roughly by the factor
\cite{Hu:1997mj}
\begin{equation}
\frac{\Delta P(k)}{P(k)} = \frac{\Delta
T(k,\tau=\tau_0)}{T(k,\tau=\tau_0)}\simeq -8
\frac{\Omega_\nu}{\Omega_m},
\end{equation}
as long as $\Omega_\nu \ll \Omega_m$. The numerical factor 8 is
derived from a numerical solution of the Boltzmann equation, but
the general structure of the equation is simple to understand. At
scales smaller than the free-streaming scale the neutrino
perturbations are washed out completely, leaving only
perturbations in the non-relativistic matter (CDM and baryons).
Therefore the {\it relative} suppression of power is proportional
to the ratio of neutrino energy density to the overall matter
density. Clearly the above relation only applies when $\Omega_\nu
\ll \Omega_m$, when $\Omega_\nu$ becomes dominant the spectrum
suppression becomes exponential as in the pure hot dark matter
model. This effect is shown for different neutrino masses in
Fig.~\ref{fig:nutrans}.

The effect of massive neutrinos on structure formation only
applies to the scales below the free-streaming length. For
neutrinos with masses of several eV the free-streaming scale is
smaller than the scales which can be probed using present CMB data
and therefore the power spectrum suppression can be seen only in
large scale structure data. On the other hand, neutrinos of sub-eV
mass behave almost like a relativistic neutrino species for CMB
considerations. The main effect of a small neutrino mass on the
CMB is that it leads to an enhanced early ISW effect. The reason
is that the ratio of radiation to matter at recombination becomes
larger because a sub-eV neutrino is still relativistic or
semi-relativistic at recombination. With the WMAP data alone it is
very difficult to constrain the neutrino mass, and to achieve a
constraint which is competitive with current experimental bounds
it is necessary to include LSS data from 2dF or SDSS. When this is
done the bound becomes very strong, somewhere in the range of
1-1.5 eV for the sum of neutrino masses, depending on assumptions
about priors
\cite{Spergel:2003cb,Hannestad:2003xv,Allen:2003pt,Tegmark:2003ud,Barger:2003vs,%
Crotty:2004gm,numass7,numass8,numass9,numass10,sth2}

The bound can be strengthened even further by including data from
the Lyman-$\alpha$ forest, measured at large redshifts. This was
done for instance in \cite{numass7,numass8}, where bounds as
strong as 0.4 eV were derived. However, the systematic errors in
extracting the matter power spectrum from the Lyman-$\alpha$ flux
power spectrum are at present not understood at a level where this
bound can be claimed to be robust.

\subsubsection{Parameter degeneracy with the dark energy equation
of state}

\begin{figure}[htb]
\begin{center}
\epsfig{file=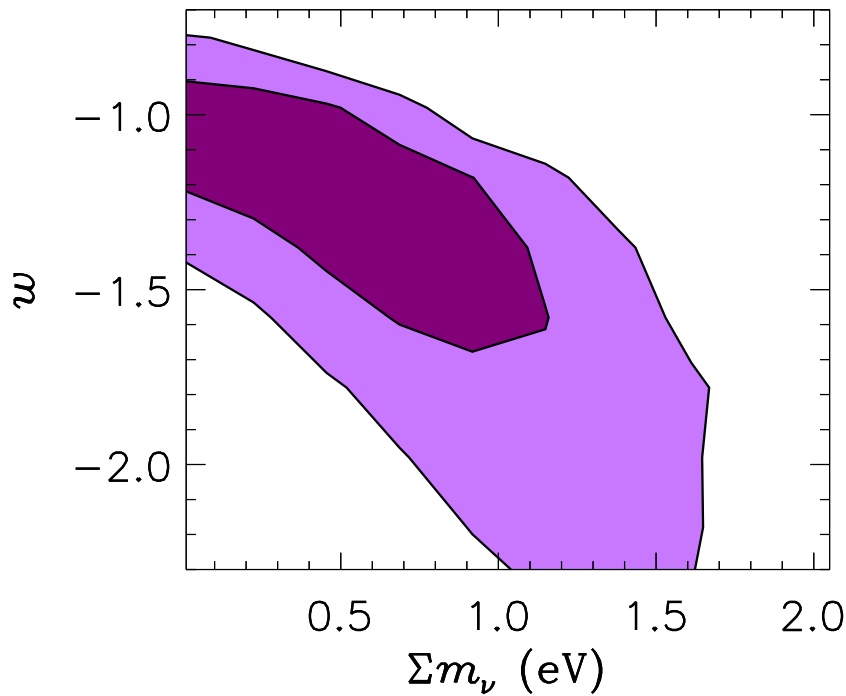,width=0.6\textwidth}
\end{center}
\caption{68\% and 95\% allowed contours as a function of neutrino
mass and dark energy equation of state using WMAP, SDSS, HST, and
SNI-a data.} \label{darkenergy}
\end{figure}

One caveat of most cosmological parameter analyses used to probe
particle physics is that they are done with the minimal
cosmological standard model. In principle there could easily be
additional parameters which are important. One such example was
described in \cite{sth2}, where it was shown that there is an
almost perfect degeneracy between the neutrino mass and the
equation of state of the dark energy, $w$. This degeneracy is
shown in Fig.~\ref{darkenergy}: An increasing $\sum m_\nu$ can be
compensated by decreasing $w$. While for low neutrino masses a
cosmological constant ($w=-1$) is allowed, for high neutrino
masses only dark energy models in the phantom regime ($w < -1$)
are allowed.

The reason for the degeneracy is that when $\Omega_\nu$ is
increased, $\Omega_m$ must be increased correspondingly in order
to produce the same power spectrum. However, when $w=-1$ an
increasing $\Omega_m$ quickly becomes incompatible with the
supernova data. This can be remedied by simultaneously decreasing
$w$ because of the well-known $\Omega_m,w$ degeneracy in the
supernova data. For the particular data set used in this case the
95\% C.L. bound changes from 0.65 eV with $w=-1$ to 1.5 eV with
free $w$.

\subsubsection{Combining measurements of $m_\nu$ and $N_\nu$.}

\begin{figure}[htbp]
\begin{center}
\epsfig{file=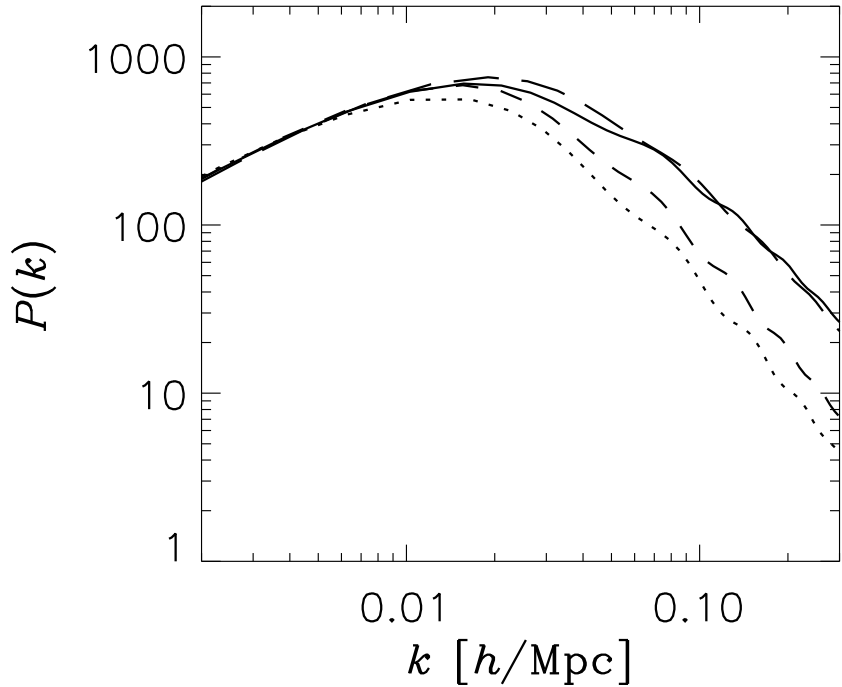,width=0.6\textwidth}
\end{center}
\bigskip
\caption{Power spectra for $\Lambda$CDM
  models with $\Omega_b = 0.05$, $\Omega = 1$, $h=0.7$, $n_s=1$, and
  $N_{\nu,{\rm massive}}=1$ and a common large-scale normalization.
  The full line is for $\Omega_\nu=0$, $\Omega_m=0.25$, $N_\nu=3$,
  dashed is for $\Omega_\nu=0.05$, $\Omega_m=0.25$, $N_\nu=3$, dotted
  is for $\Omega_\nu=0.05$, $\Omega_m=0.25$, $N_\nu=8$, and
  long-dashed is for $\Omega_\nu=0.05$, $\Omega_m=0.35$, $N_\nu=8$
  (from \protect\cite{hr04}).} \label{fig:pspec}
\end{figure}

The limits on neutrino masses discussed above apply only for
neutrinos within the standard model, i.e.\ three light neutrinos
with degenerate masses (if the sum is close to the upper bound).
However, if there are additional neutrino species sharing the
mass, or neutrinos have significant chemical potentials this bound
is changed. Models with massive neutrinos have suppressed power at
small scale, with suppression proportional to
$\Omega_\nu/\Omega_m$. Adding relativistic energy further
suppresses power at scales smaller than the horizon at
matter-radiation equality. For the {\it same} matter density such
a model would therefore be even more incompatible with data.
However, if the matter density is increased together with $m_\nu$,
and $N_\nu$, excellent fits to the data can be obtained. This
effect is shown in Fig.~\ref{fig:pspec}.

\begin{figure}[htbp]
\begin{center}
\epsfig{file=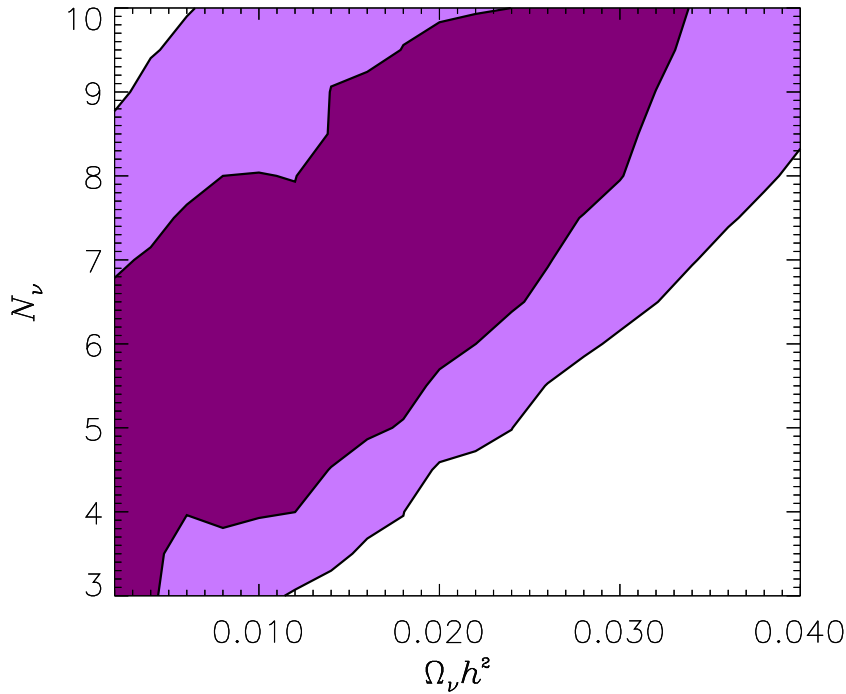,width=0.6\textwidth}
\end{center}
\bigskip
\caption{Likelihood contours (68\% and 95\%) for the case of
$N_\nu$ neutrinos with equal masses, calculated from WMAP and 2dF
data (from \protect\cite{hr04}).} \label{fig:mnunnu}
\end{figure}

The effect on likelihood contours for $(\Omega_\nu,N_\nu)$ can be
seen in Fig.~\ref{fig:mnunnu} which is for the case where $N_\nu$
species the total mass equally.

A thorough discussion of these models can be found in
Refs.~\cite{hr04,Crotty:2004gm}.

\subsubsection{Future neutrino mass measurements}

The present bound on the sum of neutrino masses is still much
larger than the mass difference, $\Delta m_{23} \sim 0.05$ eV
\cite{Fogli:2003th,Maltoni:2003da}, measured by atmospheric
neutrino observatories and K2K . This means that if the sum of
neutrino masses is anywhere close to saturating the bound then
neutrino masses must the almost degenerate. The question is
whether in the future it will be possible to measure masses which
are of the order $\Delta m_{23}$, i.e.\ whether it can determined
if neutrino masses are hierarchical.

By combining future CMB data from the Planck satellite with a
galaxy survey like the SDSS it has been estimated that neutrino
masses as low as about 0.1 eV can be detected
\cite{Hannestad:2002cn,pastor04}. Another possibility is to use
weak lensing of the CMB as a probe of neutrino mass. In this case
it seems likely that a sensitivity below 0.1 eV can also be
reached with CMB alone \cite{Kaplinghat:2003bh}. Possibly the
sensitivity could be increased even further with future galaxy
cluster surveys \cite{Wang:2005vr}.

As noted in Ref.~\cite{pastor04} the exact value of the
sensitivity at this level depends both on whether the hierarchy is
normal or inverted, and the exact value of the mass splittings.

\subsection{Neutrino warm dark matter}

While CDM is defined as consisting of non-interacting particles
which have essentially no free-streaming on any astronomically
relevant scale, and HDM is defined by consisting of particles
which become non-relativistic around matter radiation equality or
later, warm dark matter is an intermediate. One of the simplest
production mechanisms for warm dark matter is active-sterile
neutrino oscillations in the early universe
\cite{Hansen:2001zv,Abazajian:2001vt,Abazajian:2001nj,Shi:1998km,%
Dodelson:1993je}.

One possible benefit of warm dark matter is that it does have some
free-streaming so that structure formation is suppressed on very
small scales. This has been proposed as an explanation for the
apparent discrepancy between observations of galaxies and
numerical CDM structure formation simulations. In general
simulations produce galaxy halos which have very steep inner
density profiles $\rho \propto r^\alpha$, where $\alpha \sim
1-1.5$, and numerous subhalos
\cite{Kazantzidis:2003hb,Ghigna:1999sn}. Neither of these
characteristics are seen in observations and the explanation for
this discrepancy remains an open question. If dark matter is warm
instead of cold, with a free-streaming scale comparable to the
size of a typical galaxy subhalo then the amount of substructure
is suppressed, and possibly the central density profile is also
flattened
\cite{Yoshida:2003rm,Haiman:2001dg,Avila-Reese:2000hg,Bode:2000gq,%
Colin:2000dn,Hannestad:2000gt,jsd} . In both cases the mass of the
dark matter particle should be around 1 keV
\cite{Hogan:2000bv,Dalcanton:2000hn}, assuming that it is
thermally produced in the early universe.

On the other hand, from measurements of the Lyman-$\alpha$ forest
flux power spectrum it has been possible to reconstruct the matter
power spectrum on relatively small scales at high redshift. This
spectrum does not show any evidence for suppression at sub-galaxy
scales and has been used to put a lower bound on the mass of warm
dark matter particles of roughly 500 eV \cite{Viel:2005qj}. An
even more severe problem lies in the fact that star formation
occurs relatively late in warm dark matter models because small
scale structure is suppressed. This may be in conflict with the
low-$l$ CMB temperature-polarization cross correlation measurement
by WMAP which indicates very early reionization and therefore also
early star formation. One recent investigation of this found warm
dark matter to be inconsistent with WMAP for masses as high as 10
keV \cite{Yoshida:2003rm}.

The case for warm dark matter therefore seems quite marginal,
although at present it is not definitively ruled out by any
observations.

\section{Discussion} 

Here I have reviewed some of the basics of neutrino cosmology with
particular emphasis on the role that neutrinos play in
cosmological structure formation. Exactly because neutrinos play
such a crucial role in cosmology it is possible to use
cosmological observations to probe fundamental properties of
neutrinos which are not easily accessible in lab experiments.
Particularly the measurement of absolute neutrino masses from CMB
and large scale structure data has received significant attention
over the past few years.

Another cornerstone of neutrino cosmology is the measurement of
the total energy density in non-electromagnetically interacting
particles. For many years Big Bang nucleosynthesis was the only
probe of relativistic energy density, but with the advent of
precision CMB and LSS data it has been possible to complement the
BBN measurement. At present the cosmic neutrino background is seen
in both BBN, CMB and LSS data at high significance.

Finally, cosmology can also be used to probe the possibility of
neutrino warm dark matter, which could be produced by
active-sterile neutrino oscillations.

In the coming years the steady stream of new observational data
will continue, and the cosmological bounds on neutrino will
improve accordingly. For instance, it has been estimated that with
data from the upcoming Planck satellite it could be possible to
measure neutrino masses as low as 0.1 eV, and possibly even lower
than that if data from future cluster surveys is used.

Neutrino cosmology is certainly an exiting field which is destined
to play a key role in neutrino physics in the coming years.

\end{document}